\documentclass[aps,prx,amsfonts,amsmath,amssymb,raggedbottom,longbibliography,reprint,superscriptaddress,citeautoscript]{revtex4-2}

\usepackage{graphicx} 
\usepackage{float}
\usepackage{color}
\usepackage{bm}
\usepackage{hyperref}
\usepackage{todonotes}
\usepackage{verbatim}
\usepackage{soul}
\usepackage{glossaries}
\usepackage{sidecap}
\usepackage{hyperref}
\usepackage{ulem}
\hypersetup{
    colorlinks=true,
    linkcolor=blue,
    filecolor=magenta,
    urlcolor=blue,
    citecolor=blue,
}

\begin{document}

\title{Orbital polarization, charge transfer, and fluorescence in reduced valence nickelates}

\author{M. R. Norman}
\email{norman@anl.gov}
\affiliation{Materials Science Division, Argonne National Laboratory, Lemont, IL 60439}

\author{A. S. Botana}
\affiliation{Department of Physics, Arizona State University, Tempe, AZ 85287}

\author{J. Karp}
\affiliation{Department of Applied Physics and Applied Math, Columbia University, New York, NY 10027}

\author{A. Hampel}
\affiliation{Center for Computational Quantum Physics, Flatiron Institute, 162 5th Avenue, New York, NY 10010}

\author{H. LaBollita}
\affiliation{Department of Physics, Arizona State University, Tempe, AZ 85287}
\affiliation{Center for Computational Quantum Physics, Flatiron Institute, 162 5th Avenue, New York, NY 10010}

\author{A. J. Millis}
\affiliation{Center for Computational Quantum Physics, Flatiron Institute, 162 5th Avenue, New York, NY 10010}
\affiliation{Department of Physics, Columbia University, New York, NY 10027}

\author{G. Fabbris}
\affiliation{Advanced Photon Source, Argonne National Laboratory, Lemont, IL 60439}

\author{Y. Shen}
\affiliation{Condensed Matter Physics and Materials Science Department, Brookhaven National Laboratory, Upton, NY 11973}

\author{M. P. M. Dean}
\affiliation{Condensed Matter Physics and Materials Science Department, Brookhaven National Laboratory, Upton, NY 11973}

\date{\today}

\begin{abstract}
This paper presents  a simple formalism for calculating X-ray absorption (XAS) and resonant inelastic x-ray scattering (RIXS) that has as input orbital-resolved density of states from a single-particle or many-body \textit{ab initio} calculation and is designed to capture itinerant-like features.  We use this formalism to calculate both the XAS and RIXS with input from DFT and DFT+DMFT for the recently studied reduced valence nickelates $R_4$Ni$_3$O$_8$ and $R$NiO$_2$ ($R$ a rare earth), and these results are then contrasted with those for the cuprate CaCuO$_2$ and the unreduced nickelate $R_4$Ni$_3$O$_{10}$.  In contrast to the unreduced $R_4$Ni$_3$O$_{10}$, the reduced valence nickelates as well as the cuprate show strong orbital polarization due to the dominance of $x^2-y^2$ orbitals for the unoccupied $3d$ states.  We also reproduce two key aspects of a recent RIXS experiment for $R_4$Ni$_3$O$_8$: (i) a charge transfer feature between $3d$ and oxygen $2p$ states whose energy we find to decrease as one goes from $R$NiO$_2$ to $R_4$Ni$_3$O$_8$ to the cuprate, and (ii) an energy-dependent polarization reversal of the fluorescence line that arises from hybridization of the unoccupied $3z^2-r^2$ states with $R$ 5d states.  We end with some implications of our results for the nature of the $3d$ electrons in reduced valence nickelates.
\end{abstract}

\maketitle

\section{Introduction}

Since the discovery of superconductivity in Sr-doped NdNiO$_2$ in 2019 \cite{Li19}, there has been an active debate about the nature of reduced valence nickelates and how they compare to the cuprates. An important question has been the placement of the materials on the charge transfer-Mott continuum defined by Zaanen, Sawatzky and Allen \cite{ZSA}. Recent studies \cite{OK}  suggest that the reduced nickelates are intermediate between the charge transfer and Mott limits, though the Mott limit is often assumed. The configuration of the Ni $3d$ states, in particular whether the $3z^2-r^2$ states are partially filled, is also of interest. These questions have been difficult to resolve experimentally and also less straightforward to treat in theory.  

X-ray absorption spectroscopy (XAS) and resonant inelastic x-ray scattering (RIXS) have been instrumental in addressing these matters, and such studies exist for $R$NiO$_2$ ($R$ a rare earth) and its hole-doped variant (with a formal $d$ count of 9-x, where x is the hole doping gotten by partially replacing $R$ by Sr), as well as for the stoichiometric reduced Ruddlesden-Popper phases $R_4$Ni$_3$O$_8$ (with a formal d count of 8.67).  Various x-ray edges have been studied.  Here, we focus on the nickel $L$ edge since it directly involves the $3d$ electrons, and specifically the $L_2$ edge since the $L_3$ edge overlaps with the La M$_4$ edge.

$d-d$ excitations, prominent in the cuprates,  dominate the RIXS in these nickelates in the range of 1 to 2 eV \cite{Hepting20,Rossi21,Yao22} but we do not focus on these local excitations here.   
Rather, the RIXS spectra above 2 eV are dominated by (i) a charge transfer feature and (ii) a fluorescence line.  It is these features we are interested in addressing in this paper. To accomplish this goal, we present a simple formalism that takes as input orbital-resolved density of states (DOS) from a single-particle or many-body \textit{ab initio} calculation.  We then calculate the RIXS and XAS spectra from the joint density of states by including polarization matrix elements of these states with the core hole.  

(i) We find that the higher energy $s$ polarized feature in the RIXS loss spectrum near zero incident energy (relative to the Fermi energy \footnote{Experimentally, this zero can be determined from the midpoint of the leading edge of the XAS}) correlates with the charge transfer energy between nickel $3d$ and oxygen $2p$ electrons, as observed in $R_4$Ni$_3$O$_8$ \cite{Yao22}.  (ii) We find that the fluorescence line (a diagonal in energy feature in the incident energy - loss energy plane) crosses over as a function of incident energy from $s$ polarization dominated (due to a strong signal from unoccupied $x^2-y^2$ electrons) to $p$ polarization dominated (due to a weaker signal from unoccupied $3z^2-r^2$ electrons), again in agreement with the recent RIXS measurement for $R_4$Ni$_3$O$_8$ \cite{Yao22}.  This is corroborated by XAS, which we also calculate.  These unoccupied $3z^2-r^2$ electrons are dispersive due to their hybridization with the $R$ $5d$ states.  We find similar behavior for CaCuO$_2$ due to mixing of Cu $3z^2-r^2$ with the unoccupied Ca $3d$ states.  We contrast this behavior with unreduced La$_4$Ni$_3$O$_{10}$ (with a formal $d$ count of 7.33), where little polarization contrast is seen in the fluorescence line due to the weak orbital polarization associated with the unoccupied $3d$-$e_g$ states.  Again, this is consistent with recent RIXS results mentioned below.

In Section II, we present our formalism, and in Section III, our results.  We end in Section IV with implications of our work.

\section{Methods}

\begin{figure}
\centering
\includegraphics[width=0.75\columnwidth]{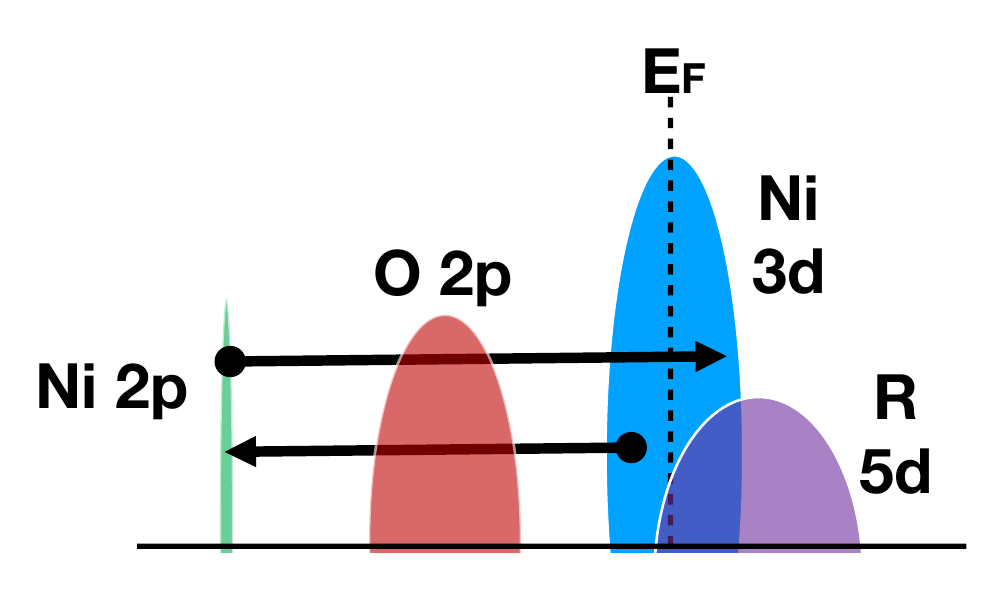}
\caption{Illustration of $L$ edge RIXS for the nickelates in the approximation used in this paper.  The upper arrow denotes excitation from the Ni $2p$ core electrons to the unoccupied Ni $3d$ states, the lower arrow denotes de-excitation from an occupied Ni $3d$ state that fills the Ni $2p$ core hole left from the first process.  The analogous XAS process only involves the upper arrow.  Here, $E_F$ denotes the Fermi energy.}
\label{fig1}
\end{figure}

Fig.~\ref{fig1} illustrates the RIXS $L$ edge process, in which an initial excitation of an electron from an atomic core Ni $2p$ state to an unoccupied Ni $3d$ state is followed by a transition from an occupied Ni $3d$ state to the hole in the Ni $2p$ shell. Following previous work \cite{Mier99}, we model this in terms of the joint density of states of empty and filled states; however, unlike Ref.~\onlinecite{Mier99}, we include polarization matrix elements of the valence/conduction electrons with the core hole. This particle-hole approximation is designed to capture more itinerant-like features such as the fluorescence line.   The resulting RIXS cross section, derived from the underlying Kramers-Heisenberg formalism \cite{Mier99,Hariki18} as shown in the Appendix, is:
\begin{eqnarray}
\tilde\sigma (E_{in},E_{out},\epsilon) & = & \sum_{i,f,\sigma,\sigma^\prime,\epsilon^\prime}
\int dE \rho_{f\sigma^\prime}(E)  \rho_{i\sigma}(E+E_{loss}) \nonumber \\
& &~~~~~~~~~~ \frac{\Gamma}{2}\frac{M_{if\sigma\sigma^\prime}(\epsilon,\epsilon^\prime)}{(E-E_{out})^2+\Gamma^2/4}
\label{eq1}
\end{eqnarray}
where $\rho_{i\sigma}$ is the unoccupied Ni $3d$ DOS for orbital $i$ and spin $\sigma$, $\rho_{f\sigma}$ is the same but for occupied states, and the energy integral is over the occupied $3d$ states such that $E + E_{loss} > E_F$ where $E_F$ is the Fermi energy, $E_{in}$ the incident energy, $E_{out}$ the outgoing energy, and $E_{loss}$ the loss energy ($E_{in}-E_{out}$), with $\Gamma$ the core-hole broadening.  Implicit in this formula is a sum over Ni sites.  Of central importance are the polarization matrix elements between the valence/conduction states and the core hole that in the dipole approximation are:
\begin{equation}
M_{if\sigma\sigma^\prime}(\epsilon,\epsilon^\prime) = |\sum_j<f\sigma^\prime|\epsilon^\prime|j><j|\epsilon|i\sigma>|^2
\label{eq2}
\end{equation}
where $\epsilon, \epsilon^\prime$ are the incoming and outgoing photon polarizations, and $j$ are the core states (two $2p_{\pm1/2}$ states for $L_2$).  For the core - conduction/valence matrix elements, we only include the angular and spin factors, as the radial factors are just an overall constant (since our results are based on band structure codes, the radial factors for all the $3d$ atomic basis orbitals associated with the muffin tins are the same). The angular integrals are Gaunt coefficients.  Assuming the surface normal $z$ is along $c$, the electric fields are:
\begin{eqnarray}
E_x & = & (Y_{1-1} - Y_{11})/\sqrt{2} \nonumber \\
E_y & = & i(Y_{1-1} + Y_{11})/\sqrt{2} \nonumber \\
E_z & = & Y_{10}
\end{eqnarray}
where $Y_{LM}$ are spherical harmonics, giving rise to the polarizations:
\begin{eqnarray}
s,s^\prime & = & \sin\phi E_x - \cos\phi E_y \nonumber \\
p & = & \cos\theta E_z + \sin\theta (\cos\phi E_x + \sin\phi E_y)  \nonumber \\
p^\prime & = & \cos\theta^\prime E_z + \sin\theta^\prime (\cos\phi E_x + \sin\phi E_y)
\end{eqnarray}
where $\theta$ is the incoming angle relative to the $xy$ plane, $\theta^\prime$ is minus the outgoing angle relative to the $xy$ plane (2$\Theta_{sc}= \theta - \theta^\prime$), and $\phi$ is the rotation angle within the $xy$ plane.
Given these expressions for $\epsilon$ and $\epsilon^\prime$, the Gaunt coefficients are then easily determined by the following angular decompositions of the core states for $L_2$:
\begin{eqnarray}
2p_{1/2} & = & (\sqrt{2}Y_{11} \downarrow - Y_{10} \uparrow)/\sqrt{3} \nonumber \\
2p_{-1/2} & = & (\sqrt{2}Y_{1-1} \uparrow - Y_{10} \downarrow)/\sqrt{3} \nonumber
\end{eqnarray}
and the $3d$ states:
\begin{eqnarray}
xy & = & i(Y_{2-2} - Y_{22})/\sqrt{2} \nonumber \\
x^2-y^2 & = & (Y_{2-2} + Y_{22})/\sqrt{2} \nonumber \\
xz & = & (Y_{2-1} - Y_{21})/\sqrt{2} \nonumber \\
yz & = & i(Y_{2-1} + Y_{21})/\sqrt{2} \nonumber \\
3z^2-r^2 & = & Y_{20} \nonumber
\end{eqnarray}
In the above formalism, we sum over outgoing polarizations $s^\prime$ and $p^\prime$, and so present results for either incoming $s$ or incoming $p$.

Similarly, we can calculate the XAS \cite{Binggeli04}:
\begin{equation}
\bar\sigma (E_{in},\epsilon) = \sum_{i,\sigma}
\int dE \frac{\Gamma}{2}\frac{\rho_{i\sigma}(E) M_{i\sigma}(\epsilon)}{(E_{in}-E)^2+\Gamma^2/4}
\end{equation}
with
\begin{equation}
M_{i\sigma}(\epsilon) = |\sum_j<i\sigma|\epsilon|j>|^2
\end{equation}
and the energy integral is over the unoccupied $3d$ states.

The above formalism has been designed for ease and generality of use and can take as input any calculation that provides an orbitally-resolved DOS.
A limitation of the formalism in its present form is that it does not take into account the interaction of the excited electron with the core hole,  and so cannot describe Raman-like features such as $d-d$ excitations; however, we observe that it can be generalized to do so if one derives the DOS from a supercell with a core hole on the absorbing site as recently demonstrated for osmates \cite{Antonov22}.

For input, we use previously published density functional theory (DFT) results for LaNiO$_2$ \cite{Antia20}, CaCuO$_2$ \cite{Antia20}, 
La$_4$Ni$_3$O$_8$ \cite{Antia16}, and La$_4$Ni$_3$O$_{10}$ \cite{Zhang20}, and dynamical mean field theory (DFT+DMFT) results for NdNiO$_2$ \cite{KarpPRX,KarpPRB}, CaCuO$_2$ \cite{KarpPRX,KarpPRB} and Pr$_4$Ni$_3$O$_8$ \cite{KarpPRB}.  In all cases, the $4f$ electrons were treated as core electrons, and so the choice of the rare earth ion is not important for the purposes of this paper.  For $\Gamma$, we assume 0.6 eV appropriate for the Ni $L_2$ edge \cite{Yao22}.
Since the inputs we use have tetragonal symmetry, the results are independent of $\phi$. DOS are in states per eV per Ni.  Although the presented RIXS and XAS cross sections are only proportional (because of the radial matrix elements), all calculations use the same normalization (and are per Ni ion).

\section{Results}

\begin{figure}
\centering
\includegraphics[width=\columnwidth]{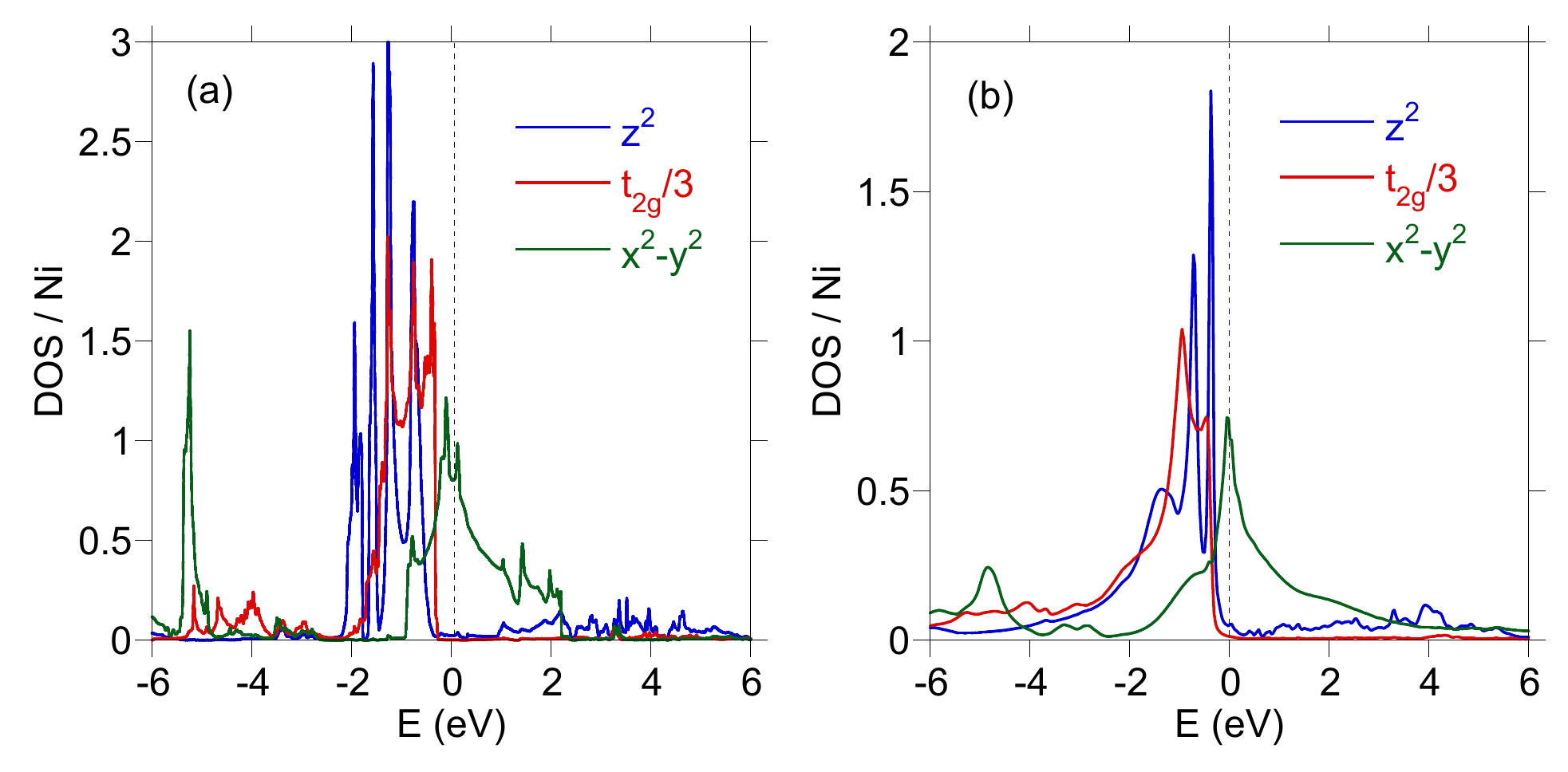}
\includegraphics[width=\columnwidth]{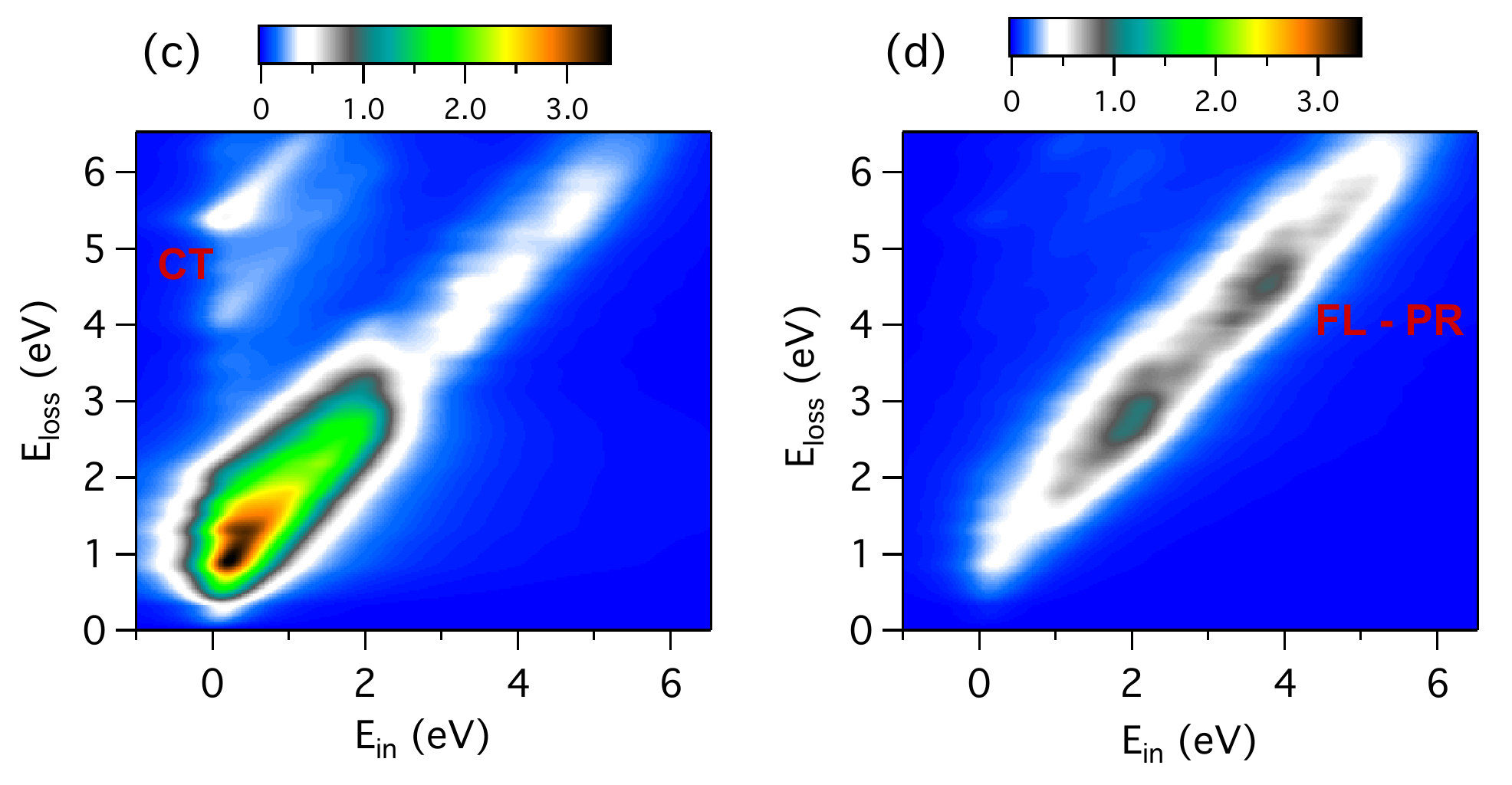}
\includegraphics[width=\columnwidth]{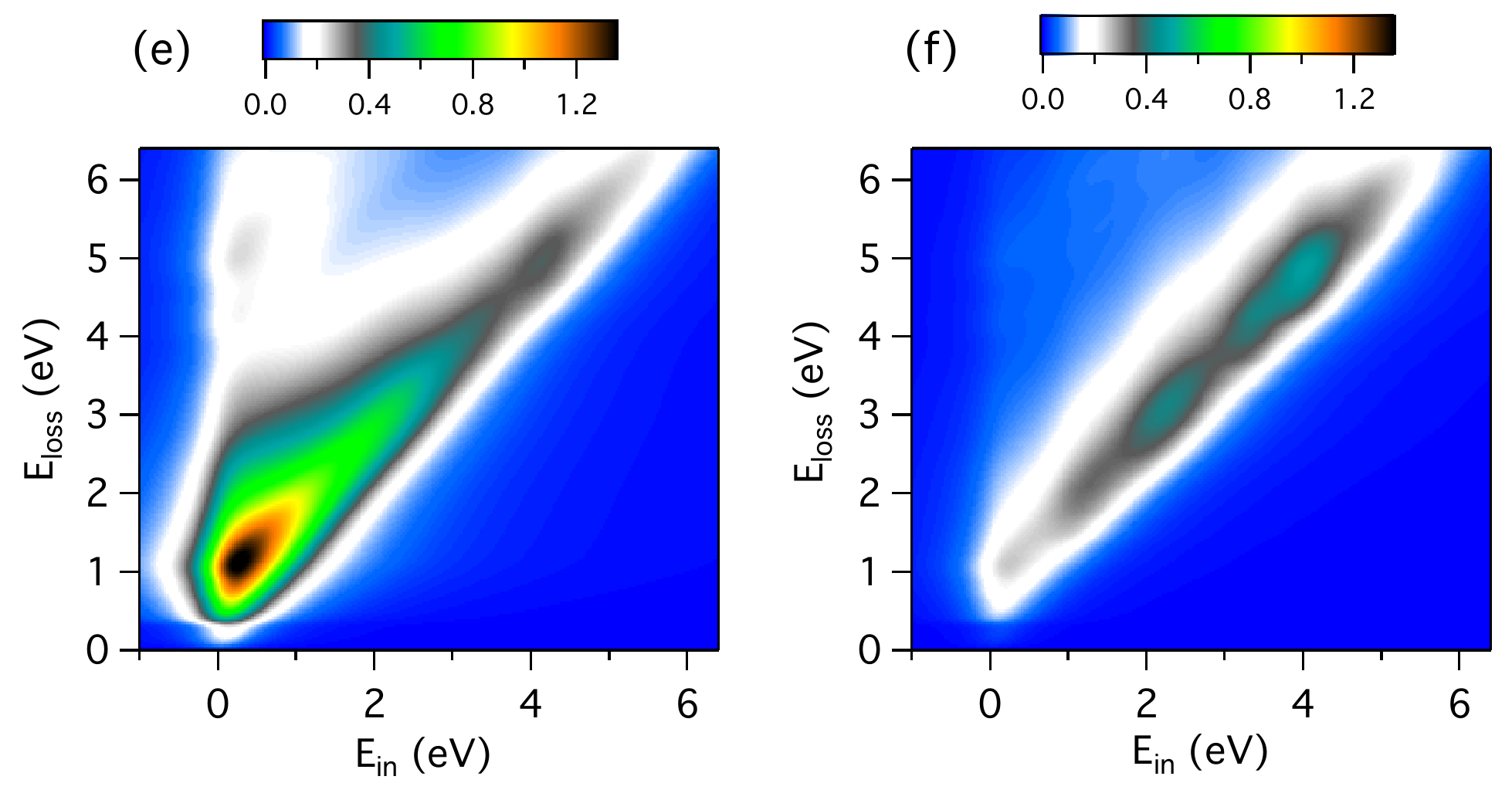}
\includegraphics[width=\columnwidth]{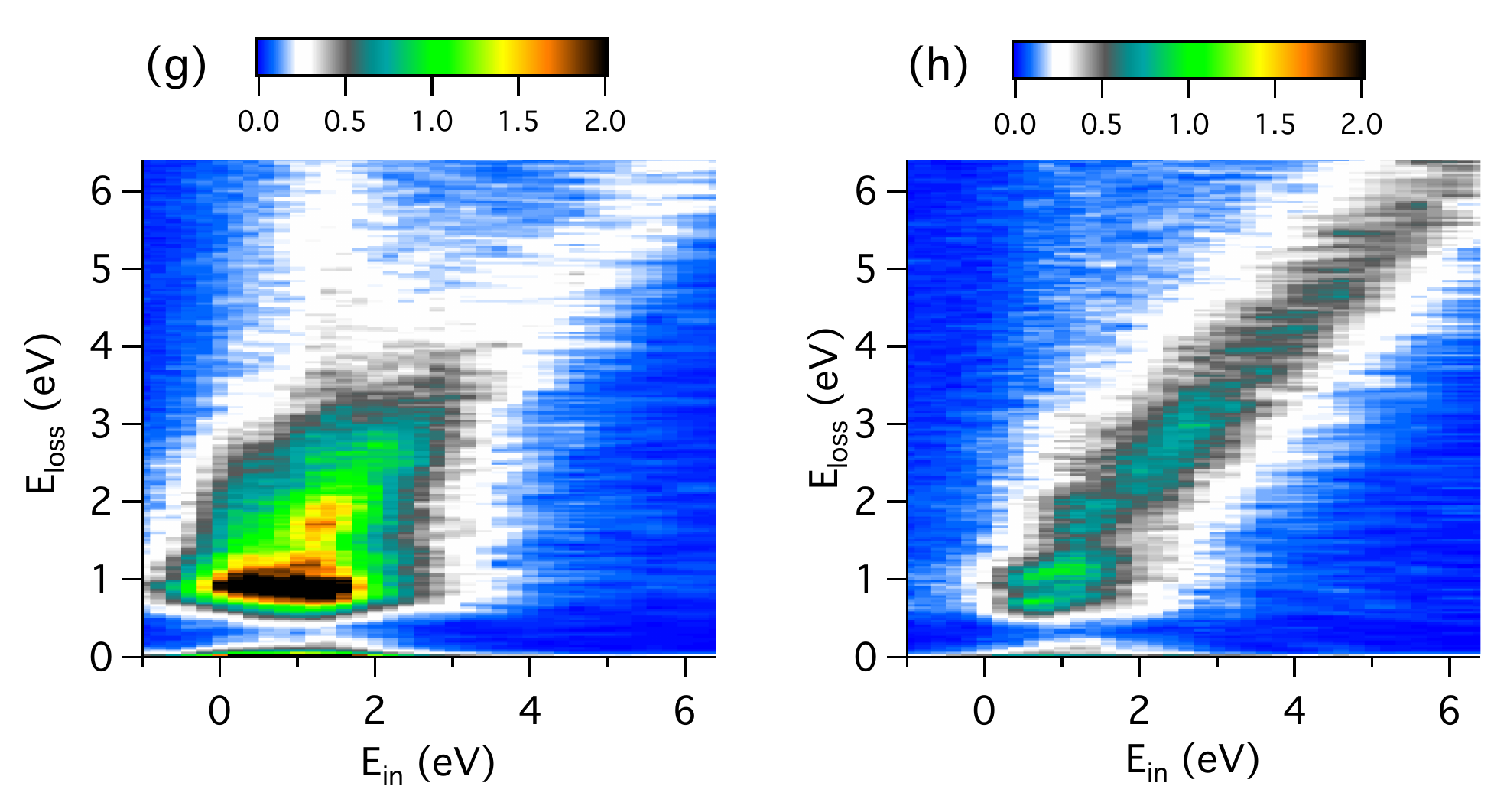}
\caption{Density of states for $R_4$Ni$_3$O$_8$ for the outer Ni planes from (a) DFT ($R$=La) and (b) DFT+DMFT ($R$=Pr). The ones for the inner Ni planes are similar. RIXS spectra for La$_4$Ni$_3$O$_8$ from DFT for $s$ polarization (c) and $p$ polarization (d).  CT denotes the charge transfer feature, FL-PR the polarization reversal of the fluorescence line.  RIXS spectra for Pr$_4$Ni$_3$O$_8$ from DFT+DMFT for $s$ polarization (e) and $p$ polarization (f).  RIXS data for La$_4$Ni$_3$O$_8$ for $s$ polarization (g) and $p$ polarization (h) from Ref.~\onlinecite{Yao22}, replotted as described in the text.  For (c-h), $\theta$=15$^\circ$ and 2$\Theta_{sc}$=153$^\circ$.}
\label{fig2}
\end{figure}

\begin{figure}
\centering
\includegraphics[width=\columnwidth]{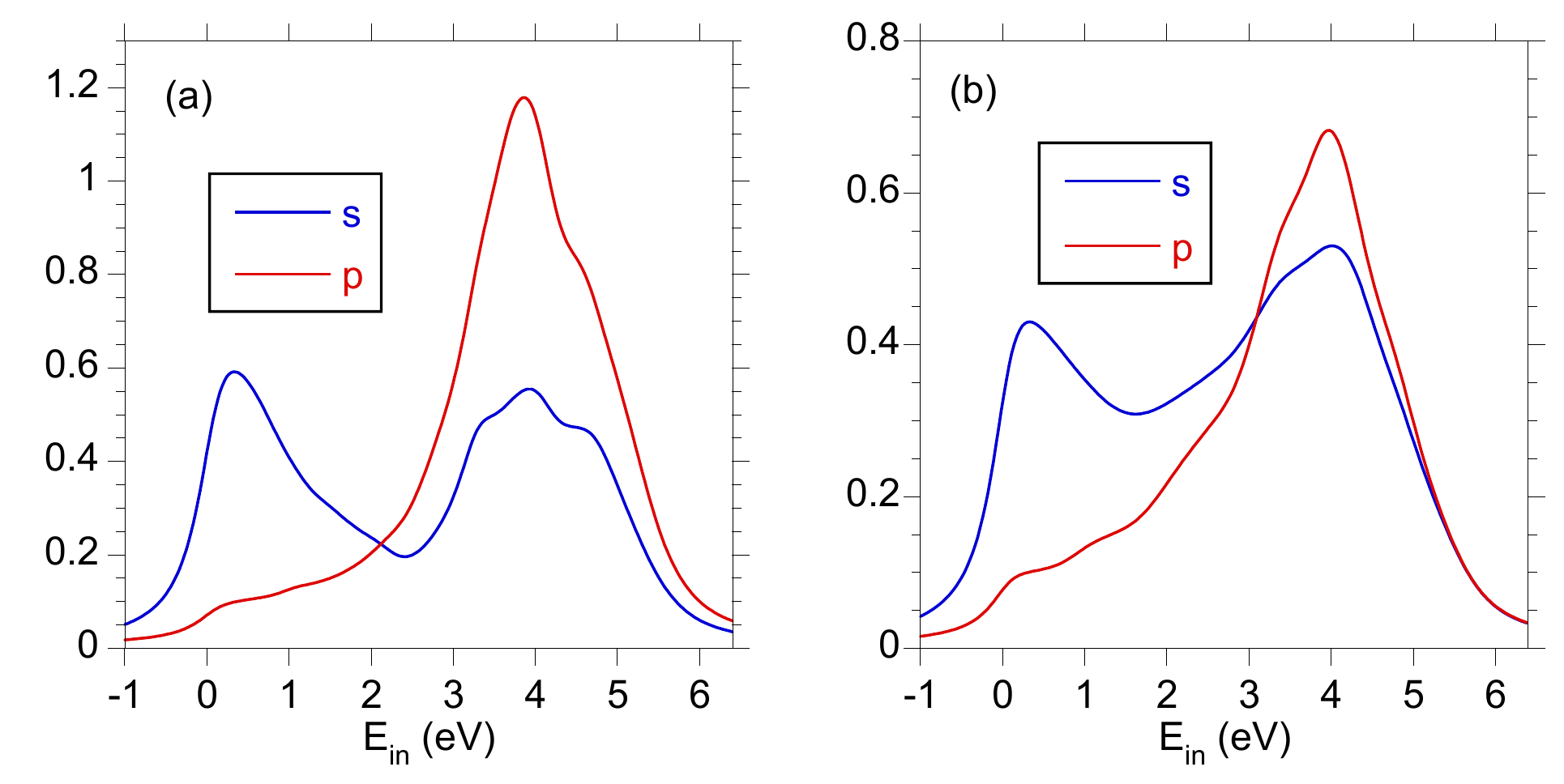}
\includegraphics[width=\columnwidth]{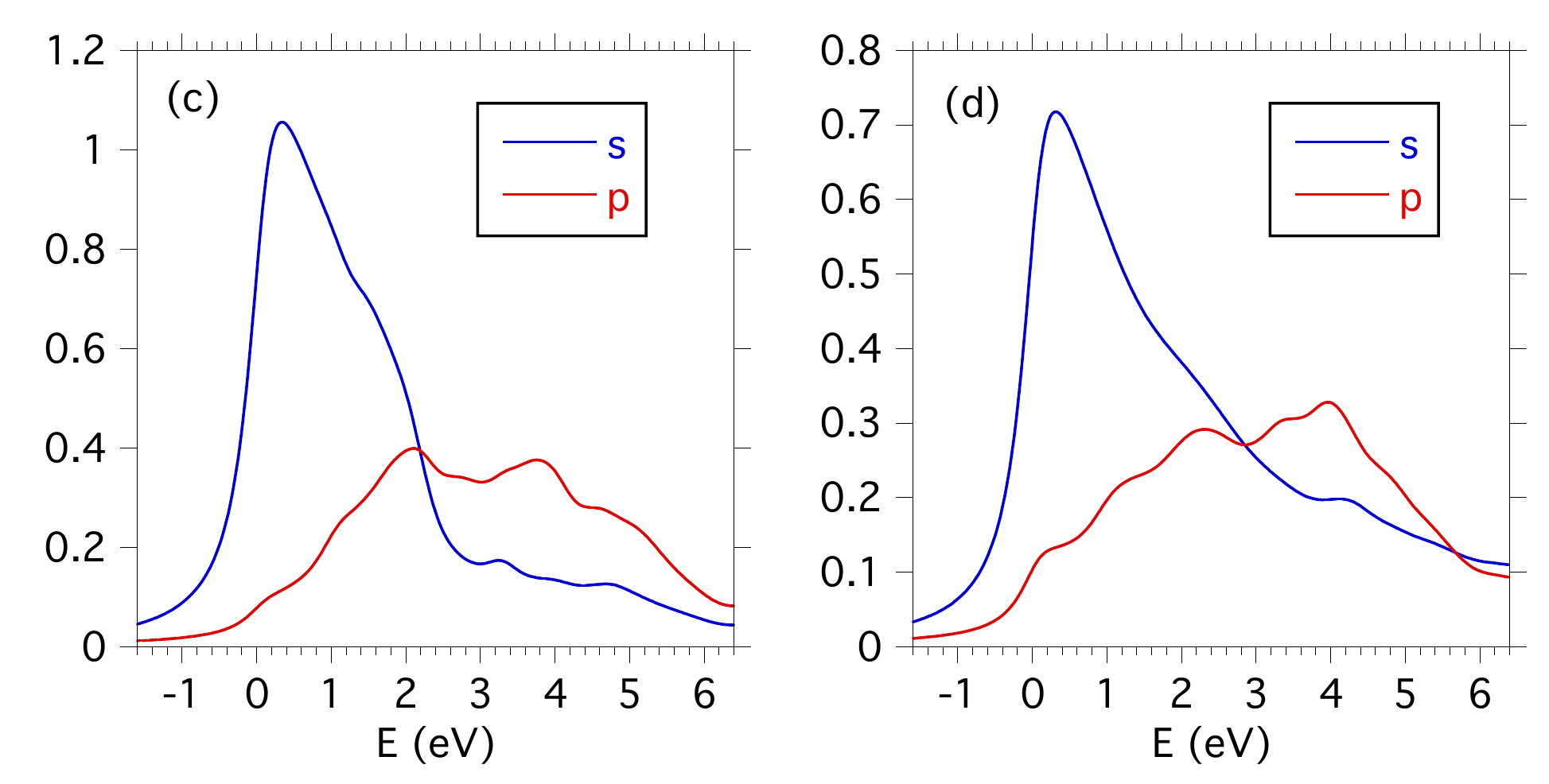}
\caption{Integration of RIXS spectra for $R_4$Ni$_3$O$_8$ from 4-6 eV loss from DFT (a, $R$=La from Fig.~\ref{fig2}(c-d)) and from DFT+DMFT (b, $R$=Pr from Fig.~\ref{fig2}(e-f)).  XAS for $R_4$Ni$_3$O$_8$ from DFT (c, $R$=La) and from DFT+DMFT (d, $R$=Pr) with $\theta$=15$^\circ$.}
\label{fig3}
\end{figure}

We first consider $R_4$Ni$_3$O$_8$.  This trilayer nickelate (one inner NiO$_2$ plane, two outer NiO$_2$ planes) is reduced from its parent Ruddlesden-Popper phase  $R_4$Ni$_3$O$_{10}$ by removal of its apical oxygens and has a formal $3d$ count of 8.67, making it analogous to overdoped cuprates.  An advantage of these 438 trilayer nickelates is that good quality bulk single crystals exist \cite{Zhang16,Zhang17}, unlike the infinite-layer 112 nickelates where thin film growth is required, enabling a wider range of experiments including high-quality RIXS data \cite{Yao22}.  

To begin our analysis, we show in Fig.~\ref{fig2}(a) the DFT-derived nickel $3d$ orbitally-resolved DOS for the outer Ni planes of La$_4$Ni$_3$O$_8$ (the ones for the inner plane are similar).  Near the Fermi level, the dominant DOS is that of the $x^2-y^2$ states that extends from about 1 eV below $E_F$ to about 2 eV above.  We also note the occupied feature about 5 eV below $E_F$ which is due to mixing between the Ni $x^2-y^2$ and the oxygen $2p\sigma$ states.  The other $3d$ states ($t_{2g}$ and $3z^2-r^2$) form sharp localized-like states below $E_F$, though $3z^2-r^2$ has a weaker unoccupied part that is broad in energy due to mixing with the $R$ $5d$ electrons. Fig.~\ref{fig2}(b) shows the same density of states, but now calculated using the DFT+DMFT method. The general features are similar; however, the $x^2-y^2$ spectrum is broadened in energy due to correlations.

\begin{figure}
\centering
\includegraphics[width=\columnwidth]{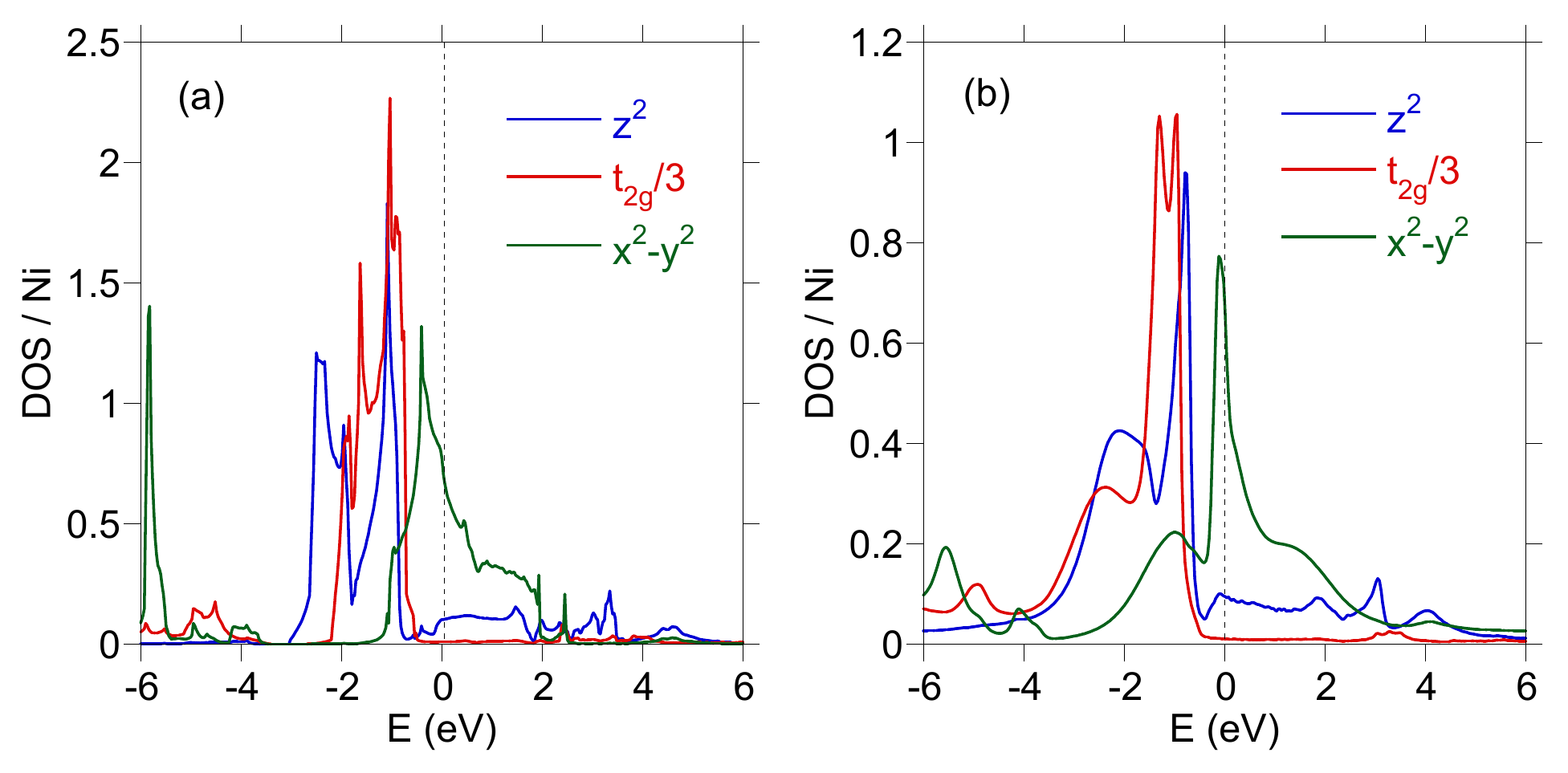}
\includegraphics[width=\columnwidth]{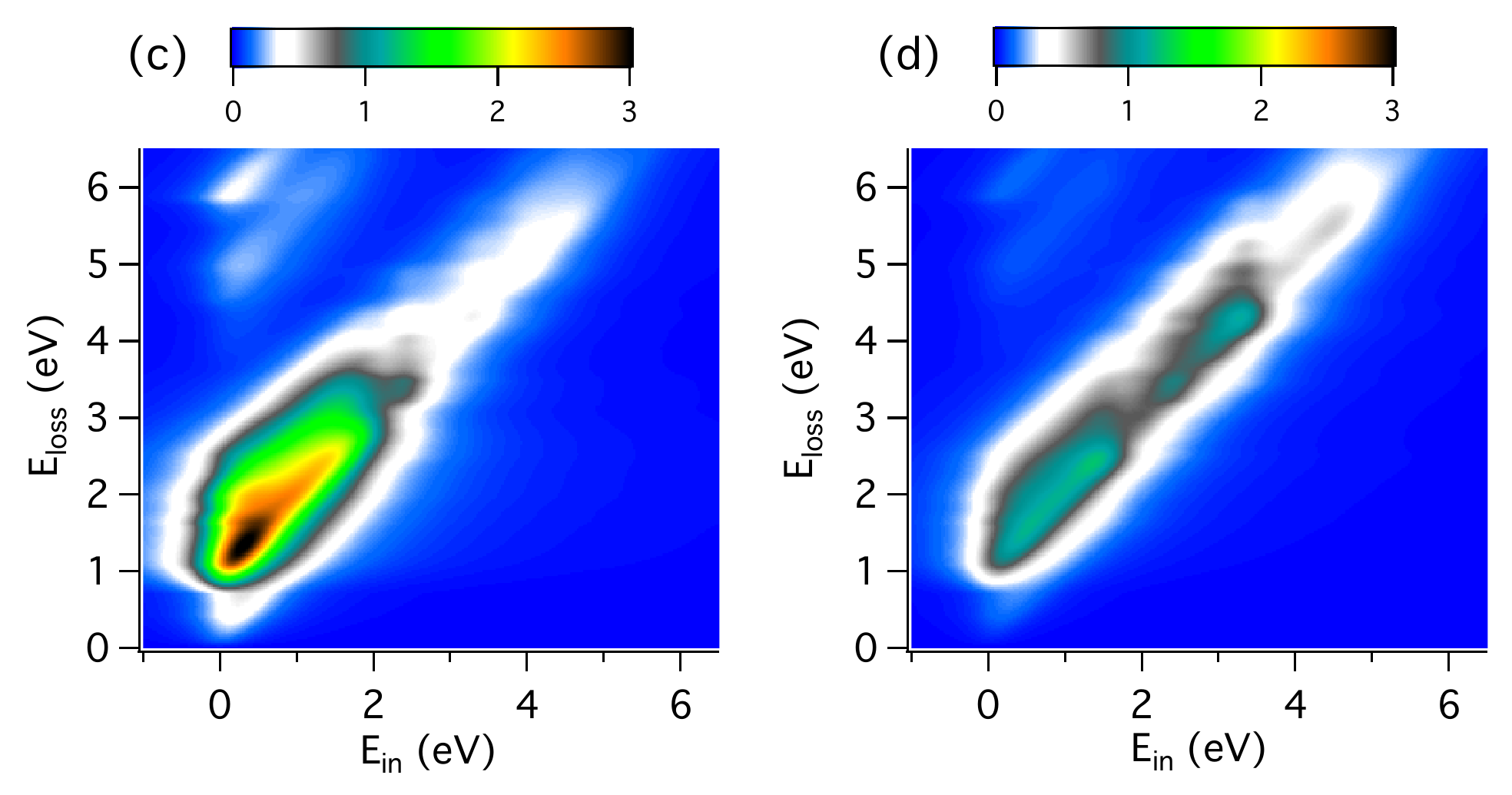}
\includegraphics[width=\columnwidth]{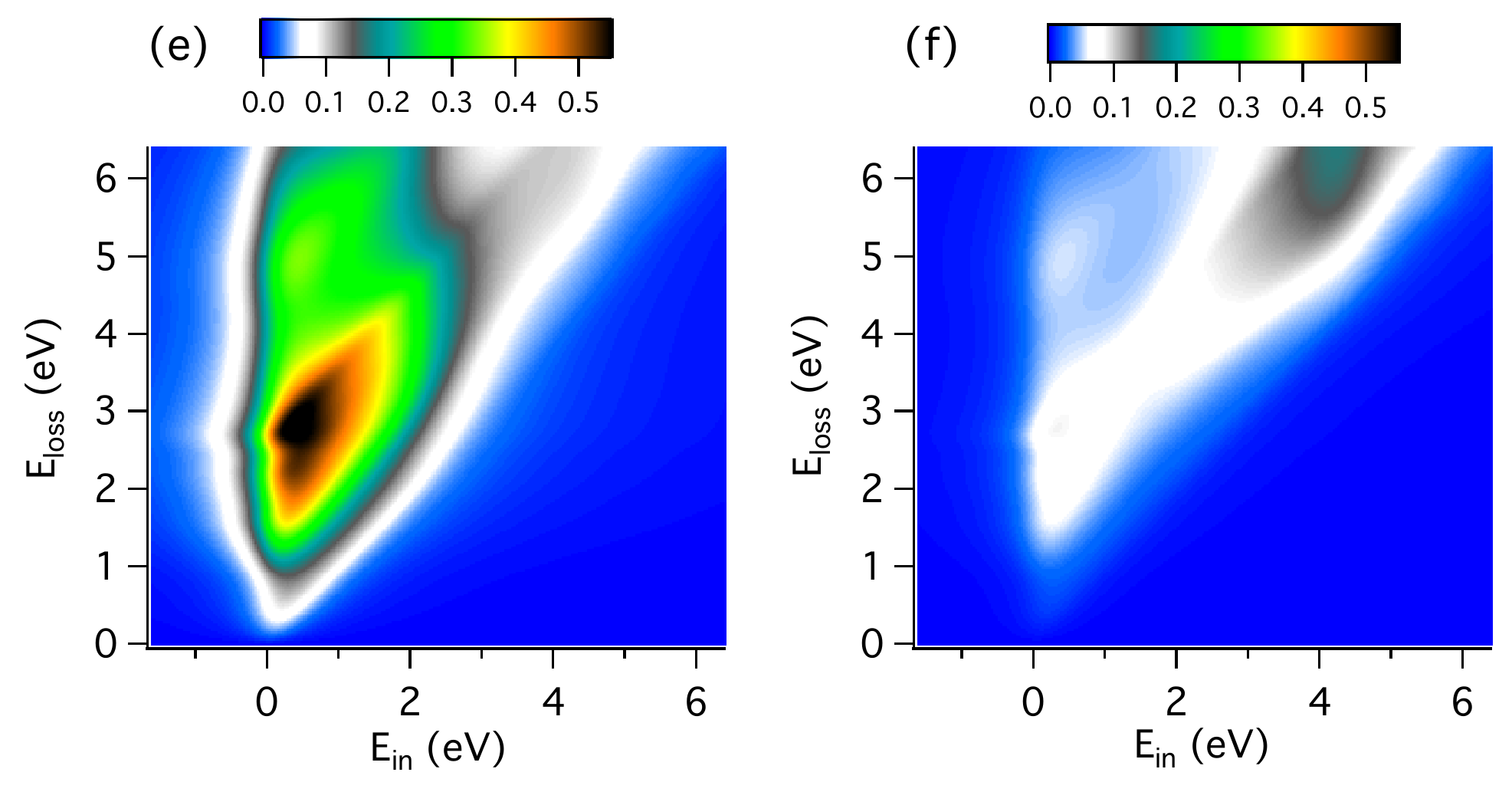}
\caption{Density of states for $R$NiO$_2$ from DFT (a, $R$=La) and DFT+DMFT (b, $R$=Nd). RIXS spectra for LaNiO$_2$ from DFT for $s$ polarization (c) and $p$ polarization (d). RIXS spectra for NdNiO$_2$ from DFT+DMFT for $s$ polarization (e) and $p$ polarization (f).  For (c-f), $\theta$=15$^\circ$ and 2$\Theta_{sc}$=153$^\circ$.}
\label{fig4}
\end{figure}

\begin{figure}
\centering
\includegraphics[width=\columnwidth]{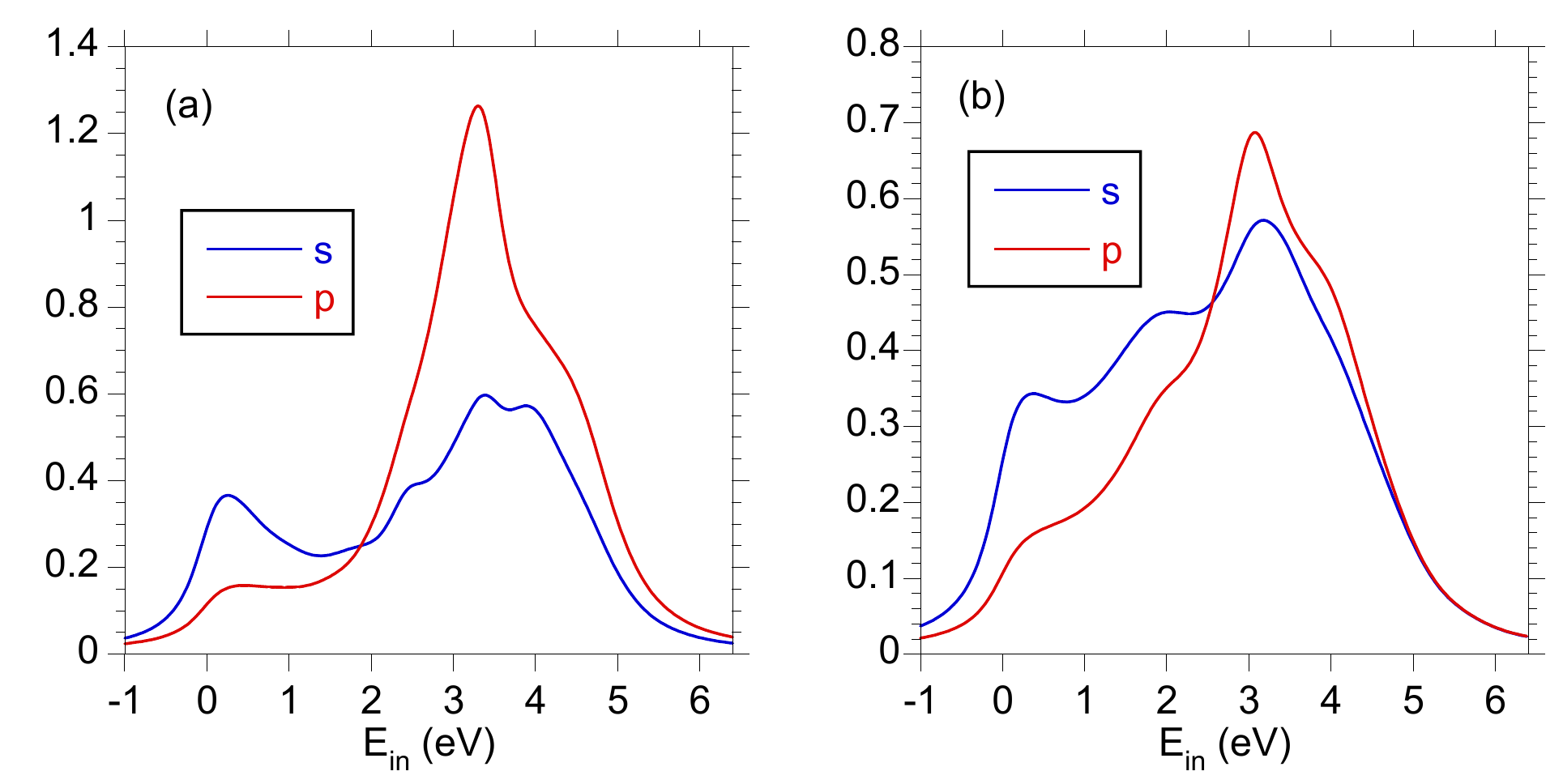}
\includegraphics[width=\columnwidth]{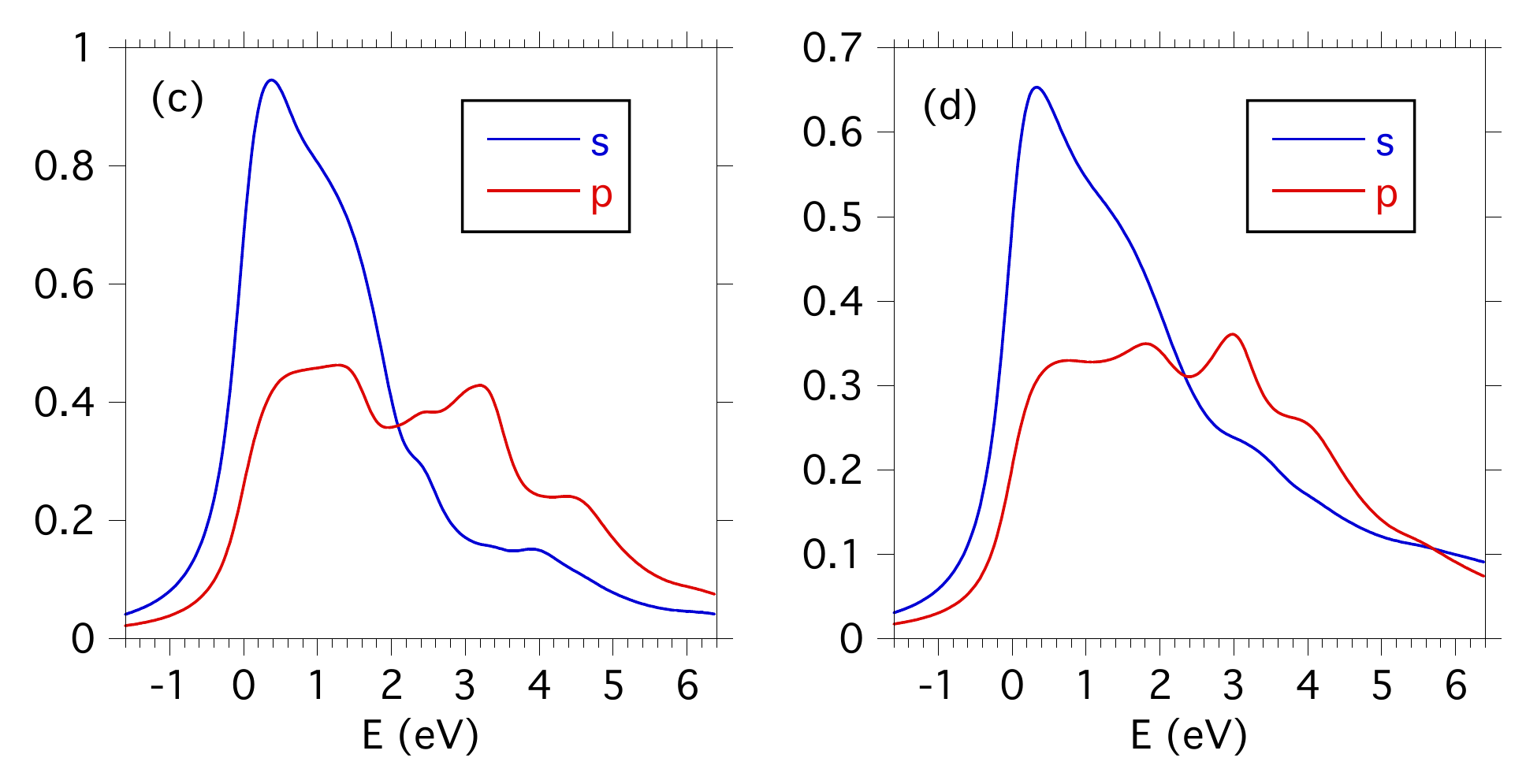}
\caption{Integration of RIXS spectra for $R$NiO$_2$ from 4-6 eV loss from DFT (a, $R$=La from Fig.~\ref{fig4}(c-d)) and from DFT+DMFT (b, $R$=Nd from Fig.~\ref{fig4}(e-f)). XAS for  $R$NiO$_2$ from DFT (c, $R$=La) and from DFT+DMFT (d, $R$=Nd) with $\theta$=15$^\circ$.}
\label{fig5}
\end{figure}

We now show how this DOS is reflected in the RIXS and XAS.
Fig.~\ref{fig2}(c-d)  shows the  RIXS spectra in the incident energy - loss energy plane at the $L_2$ edge for La$_4$Ni$_3$O$_8$ calculated using the DFT density of states shown in Fig.~\ref{fig2}(a).  The scattering angle (2$\Theta_{sc}$=153$^\circ$) was chosen to agree with that of Ref.~\onlinecite{Yao22}.  The left plot is for $s$ polarization, the right plot for $p$ polarization.  The most prominent feature in these plots is a diagonal feature which is the fluorescence line, with this diagonal behavior typical of itinerant-like states ($d-d$ excitations, on the other hand, occur at a specific loss energy).  At energies below 3 eV or so, the line is dominantly $s$ polarized, for energies above this, it is dominantly $p$ polarized.
This is seen in experiment \cite{Yao22} as shown in Fig.~\ref{fig2}(g-h) \footnote{To compare experiment to theory, we determined the experimental zero of the incident energy axis by integrating the RIXS spectra over loss energy, which is an approximation to the XAS, and then fit the leading edge to a Fermi function. To compare the intensity scales, we based the experimental scale on the non $d-d$ excitation part of the spectra, since the $d-d$ excitations are not present in the theory.}.
Given that the unoccupied $3d$ states are either $x^2-y^2$ or $3z^2-r^2$ (Fig.~\ref{fig2}(a)), we redid the calculations restricting the initial states $i$ in Eq.~\ref{eq1} to either one of these orbitals.  As expected, the $s$ signal arises from the $x^2-y^2$ states, the $p$ signal from the $3z^2-r^2$ states.  The stronger overall intensity of the $s$ signal is due to the dominance of  $x^2-y^2$ in the near $E_F$ region.  The contribution from unoccupied $3z^2-r^2$ states is more spread out in energy, a reflection of their hybridization with the dispersive $R$ $5d$ states.  This is evident as well from previous RIXS simulations of $R$NiO$_2$ where the fluorescence line was suppressed by artificially turning off the Ni $3d$-$R$ $5d$ hybridization \cite{Higashi21}. The feature near zero incident energy and $E_{loss}$ around 4-6 eV is $s$ polarized and is due to charge transfer excitations between $x^2-y^2$ and oxygen $2p\sigma$ states as mentioned before in connection with the DOS in Fig.~\ref{fig2}(a). This feature is also seen in experiment \cite{Yao22} as shown in Fig.~\ref{fig2}(g-h).

Fig.~\ref{fig2}(e-f) shows the $s$ and $p$ polarized spectra computed using the DFT+DMFT density of states. The features are similar to that obtained in the DFT calculation; the  quantitative differences  can be traced to the renormalization of the $3d$ states due to correlations for DMFT (Fig.~\ref{fig2}(b)).  This is particularly seen for the charge transfer feature  near zero incident energy and $E_{loss}$ larger than 3 eV.
In fact, these DMFT results more closely resemble the experimental data as can be seen by comparing Fig.~\ref{fig2}(e-f) to Fig.~\ref{fig2}(g-h).

The DFT and DMFT calculations only partially succeed in mirroring the data from Fig.~\ref{fig2}(g-h) in the range of 1-2 eV loss. This is because our formalism is not designed to capture the Raman-like $d-d$ features that arise from the tendency of the core hole potential well on the Ni sites in the intermediate RIXS state to localize the $3d$ states.  We refer to Ref.~\onlinecite{Yao22} for a detailed discussion of cluster calculations that do account for these $d-d$ excitations.

To investigate further, similar to Ref.~\onlinecite{Yao22}, we integrate the RIXS spectra from 4 eV to 6 eV in energy loss and plot as a function of incident energy in Fig.~\ref{fig3}(a) (DFT) and (b) (DFT+DMFT). For both calculations, one sees that that the lower incident energy part is dominantly $s$ polarized (the charge transfer feature), with the higher incident energy part strongly (DFT) or more modestly (DFT+DMFT) $p$ polarized (the fluorescence line), as in the experiment (Fig.~2d of Ref.~\onlinecite{Yao22}).
 The basic message of Figs.~\ref{fig2}-\ref{fig3} is also reflected in the XAS (Fig.~\ref{fig3}(c-d)) which again reflect a strong $s$ polarization at low energies and a more moderate $p$ polarization at higher energies.  The strong orbital polarization of the XAS, similar to that of the cuprates, has already been remarked on in previous work \cite{Zhang17}.
 
\begin{figure}
\centering
\includegraphics[width=\columnwidth]{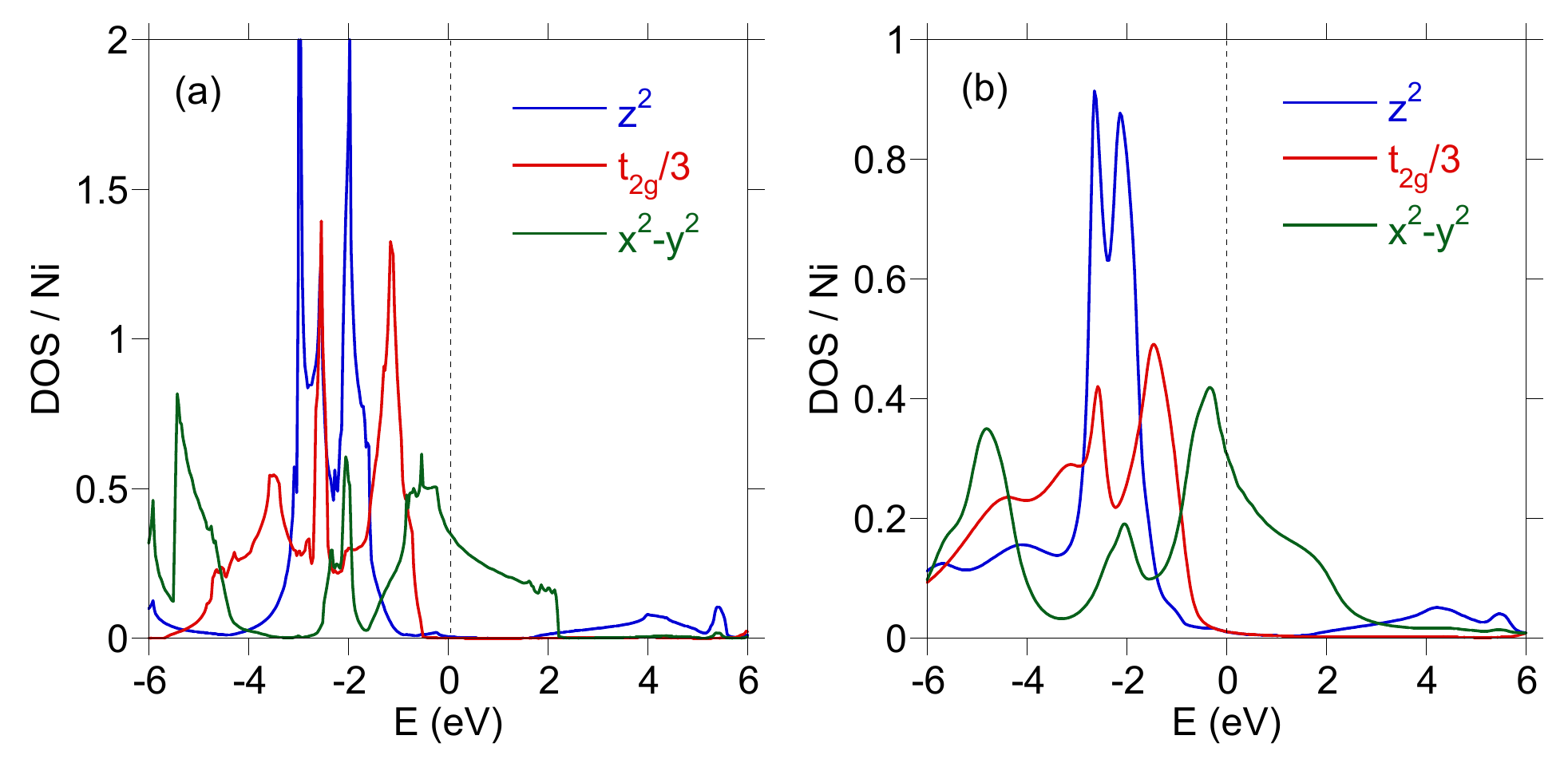}
\includegraphics[width=\columnwidth]{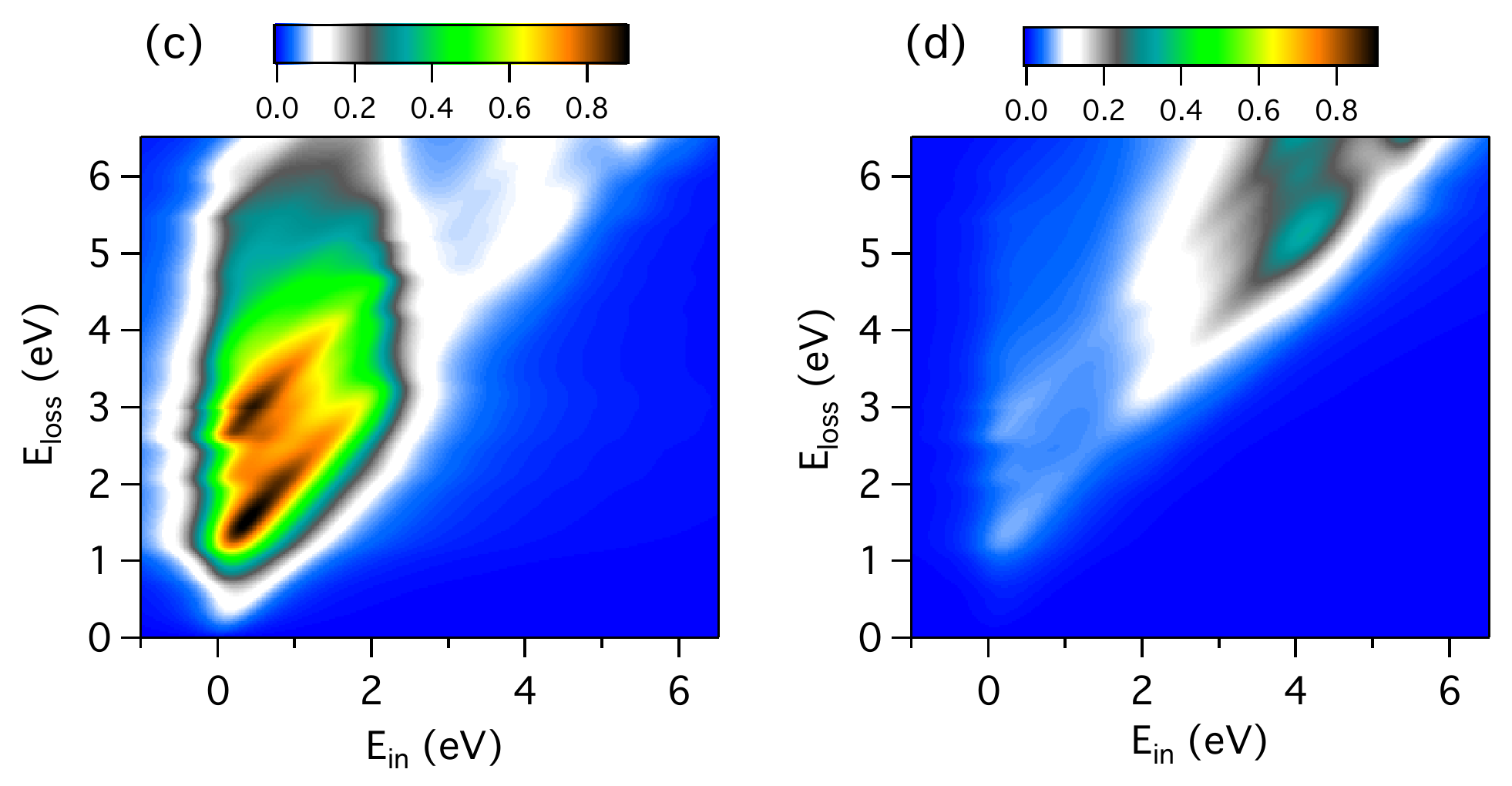}
\includegraphics[width=\columnwidth]{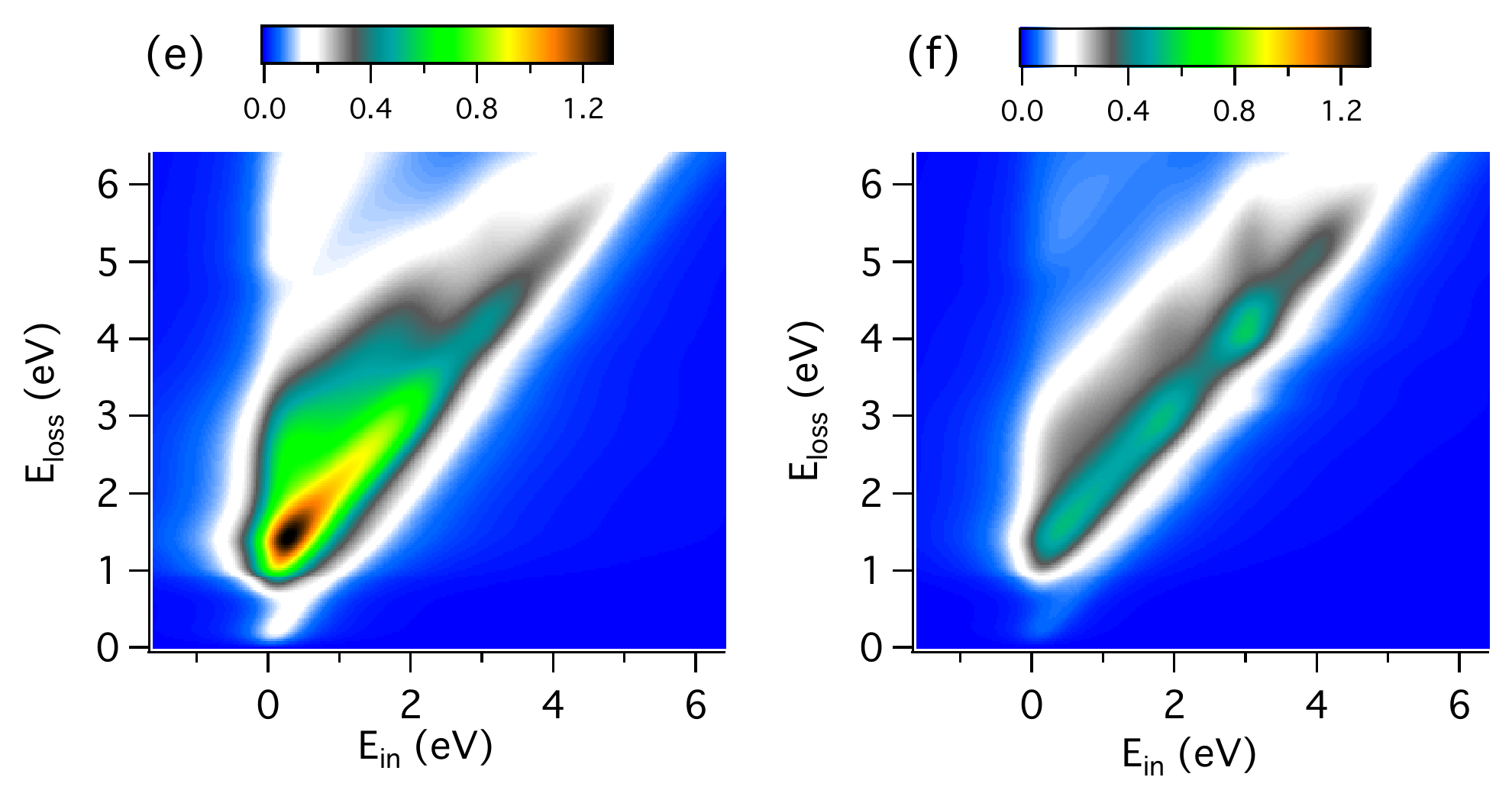}
\caption{Density of states for CaCuO$_2$ from DFT (a) and DFT+DMFT (b). RIXS spectra for CuCuO$_2$ from DFT for $s$ polarization (c) and $p$ polarization (d). RIXS spectra for CaCuO$_2$ from DFT+DMFT for $s$ polarization (e) and $p$ polarization (f). For (c-f), $\theta$=15$^\circ$ and 2$\Theta_{sc}$=153$^\circ$.}
\label{fig6}
\end{figure}

\begin{figure}
\centering
\includegraphics[width=\columnwidth]{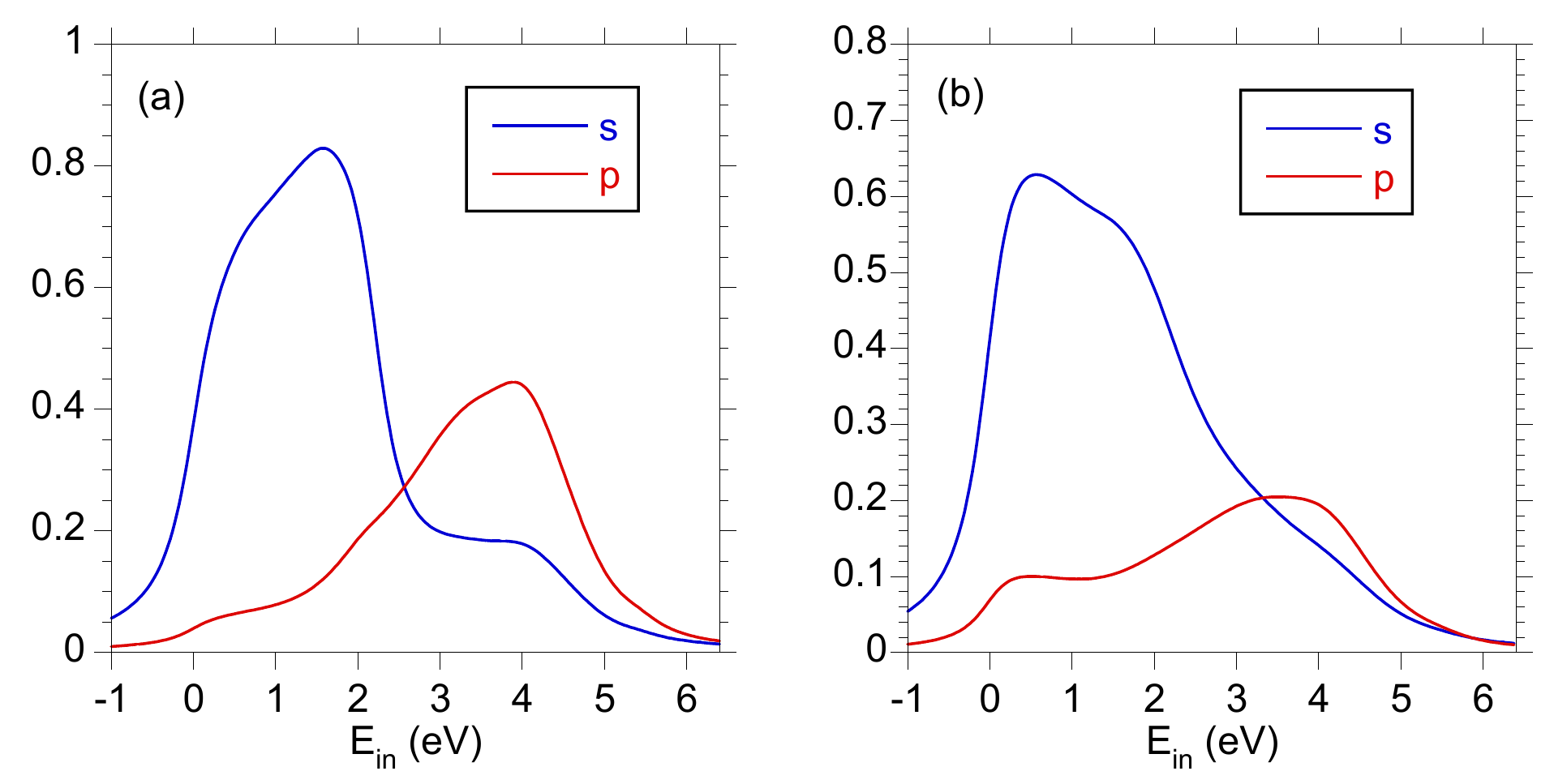}
\includegraphics[width=\columnwidth]{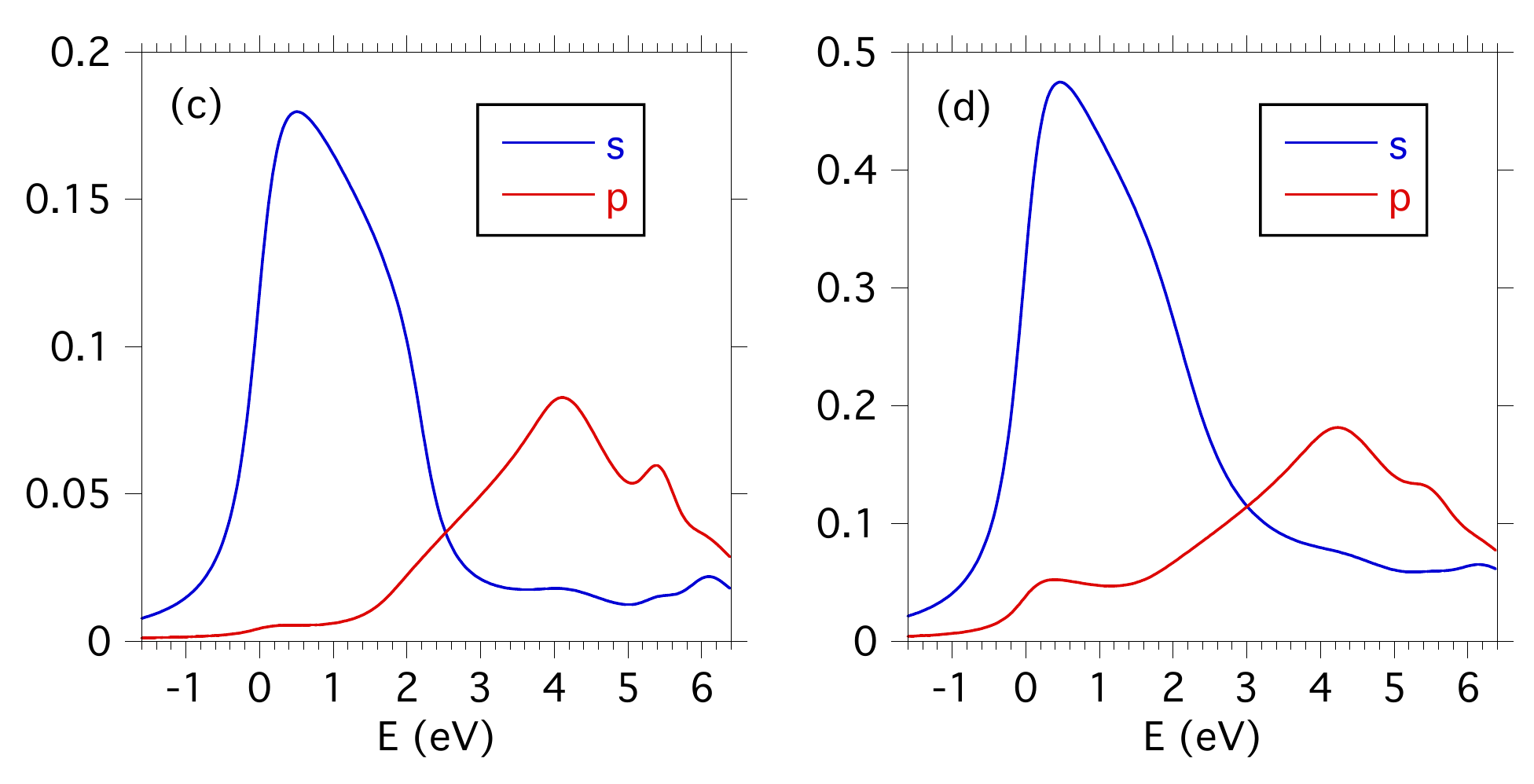}
\caption{Integration of RIXS spectra for CaCuO$_2$ from 4-6 eV loss from DFT (a, from Fig.~\ref{fig6}(c-d)) and from DFT+DMFT (b, from Fig.~\ref{fig6}(e-f)). XAS for CaCuO$_2$ from DFT (c) and from DFT+DMFT (d) with $\theta$=15$^\circ$.}
\label{fig7}
\end{figure}

Similar behavior is predicted for (undoped) $R$NiO$_2$ (Figs.~\ref{fig4}-\ref{fig5}), though with more differences for the DFT+DMFT RIXS spectra.  The fluorescence line is also quite prominent in the RIXS data \cite{Hepting20,Rossi21}, though the charge transfer feature is not so evident (though the more recent data of Ref.~\onlinecite{Rossi21} only extended up to a 4 eV loss).  We can contrast this with the cuprate CaCuO$_2$ (Figs.~\ref{fig6}-\ref{fig7}).  Although the behavior is as well similar, there are some significant differences due to the fact that the Ca $3d$ states (the analogues of the $R$ 5d states) are pushed upwards in energy by about 2 eV and the charge-transfer energy for the cuprates is about 2 eV lower.  It is the latter that makes the cuprates more charge transfer-like (doped holes sitting mostly on the oxygen sites but with appreciable nickel character) as compared to the more Mott-like behavior of the reduced valence nickelates (doped holes sitting mostly on the nickel sites but with appreciable oxygen character).  Unfortunately, almost all RIXS data on cuprates concentrate on lower energy features, though both the charge transfer and fluorescence line has been observed in Ca$_2$CuO$_2$Cl$_2$ \cite{Lebert17} \footnote{For cuprates, the $d-d$ excitations are so dominant, the charge transfer and fluorescence line features in Ref.~\onlinecite{Lebert17} required a log intensity scale to become apparent.}. Comparing all of the figures, it becomes evident that $R_4$Ni$_3$O$_8$ is intermediate in behavior between $R$NiO$_2$ and CaCuO$_2$.  This is a reflection of their charge transfer energies, that is smallest for the cuprate and largest for the undoped infinite-layer nickelate, with the difference between $R$NiO$_2$ and $R_4$Ni$_3$O$_8$ mostly due to their difference in doping.

\begin{figure}
\centering
\includegraphics[width=\columnwidth]{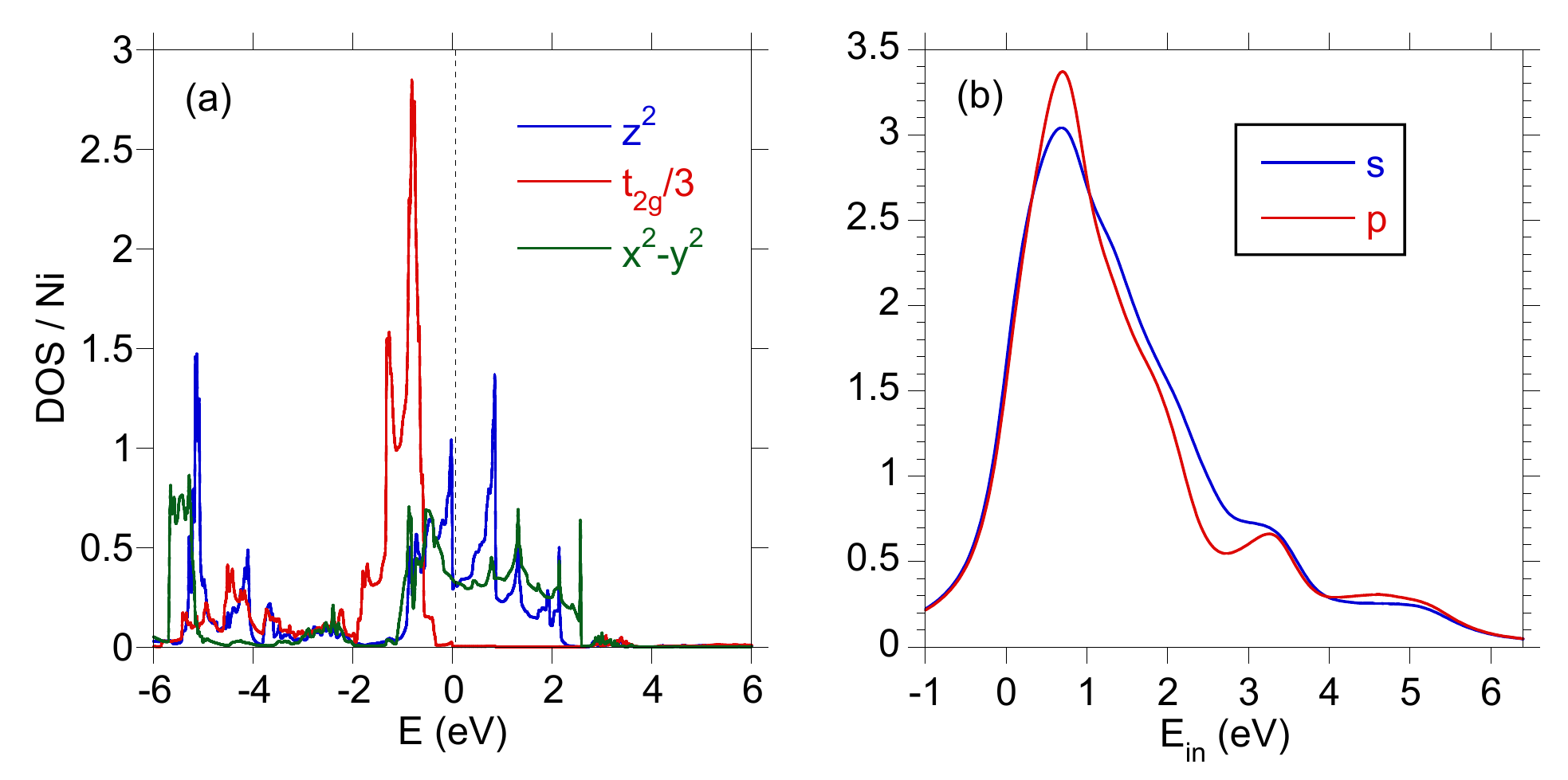}
\includegraphics[width=\columnwidth]{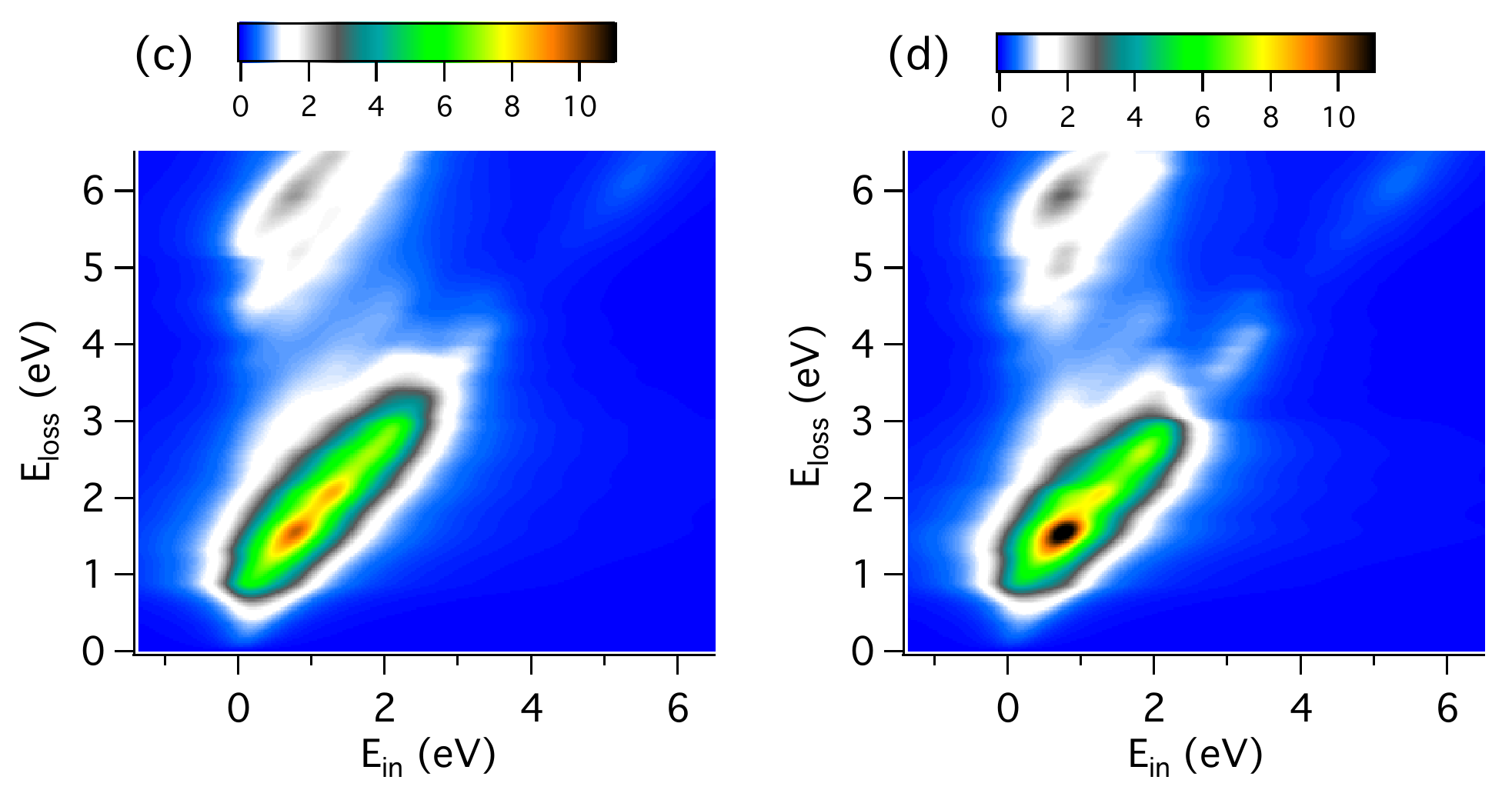}
\caption{Density of states for La$_4$Ni$_3$O$_{10}$ from DFT (a). Integration of RIXS spectra for La$_4$Ni$_3$O$_{10}$ from 4-6 eV loss from DFT (b). RIXS spectra for La$_4$Ni$_3$O$_{10}$ from DFT for $s$ polarization (c) and $p$ polarization (d) with $\theta$=15$^\circ$ and 2$\Theta_{sc}$=153$^\circ$.}
\label{fig8}
\end{figure}

The above can in turn be contrasted with results on $R_4$Ni$_3$O$_{10}$ shown in Fig.~\ref{fig8} \footnote{Although La$_4$Ni$_3$O$_{10}$ is orthorhombic, to ease comparison to the other cases, our DFT input was generated from an optimized I4/mmm structure.}. These results indicate little orbital polarization (the $s$ and $p$ signals are very similar).  For XAS, this has been previously reported \cite{Zhang17}, but RIXS intensity plots in the incident energy - energy loss plane verify this \cite{Gilberto23}.  This is because the formal $3d$ count of these unreduced nickelates is 7.33, meaning the unoccupied $3d$ states have both strong $x^2-y^2$ and $3z^2-r^2$ character. We further illustrate this by showing in Table \ref{table1} the unoccupied spectral weight for $x^2-y^2$ and $3z^2-r^2$ orbitals from the DOS plots as a representation of the orbital polarization.  Consistent with the trends discussed in this Section, the $3z^2-r^2$ unoccupied weight increases as one goes from CaCuO$_2$ to $R_4$Ni$_3$O$_8$ to $R$NiO$_2$ to $R_4$Ni$_3$O$_{10}$.

\begin{table}
\caption{Unoccupied spectral weight for $x^2-y^2$ and $3z^2-r^2$ orbitals from the 
density of states plots shown in the paper.  The first column for each orbital are the DFT results, the second column the DFT+DMFT ones. For the second row, $R$=La for DFT, $R$=Pr for DMFT.  For the third row, $R$=La for DFT, $R$=Nd for DMFT.}
\begin{ruledtabular}
\begin{tabular}{lcccc}
 & $3z^2-r^2$ &  & $x^2-y^2$ & \\
\hline
CaCuO$_2$ & 0.20 & 0.16 & 0.52 & 0.51 \\
$R_4$Ni$_3$O$_8$ & 0.36 & 0.34 & 0.92 & 0.93 \\
$R$NiO$_2$ & 0.42 & 0.40 & 0.77 & 0.83 \\
La$_4$Ni$_3$O$_{10}$ & 0.76 & 0.84 & & \\
\end{tabular}
\end{ruledtabular}
\label{table1}
\end{table}

\section{Summary}

In this paper, we presented a simple formalism for calculating RIXS fluorescence lines and other related features that are of an itinerant-like nature, and used this to address RIXS data on reduced valence nickelates.  In these simulations, we identify charge-transfer features near zero incident energy, and a polarization reversal of the fluorescence line as a function of incident energy, that match recent RIXS results for $R_4$Ni$_3$O$_8$ \cite{Yao22}.  We also presented analogous results for $R$NiO$_2$, CaCuO$_2$, and $R_4$Ni$_3$O$_{10}$.

Our results, in particular the charge-transfer feature and the dispersive fluorescence line, are a reflection of hybridization of the Ni $3d$ electrons with the oxygen $2p$ ones \cite{Harry}, with an additional contribution from unoccupied $3z^2-r^2$ states due to their hybridization with the $R$ $5d$ states.  The latter plays an important role in certain theories of reduced valence nickelates \cite{Foyevtsova22}.  Moreover, the strong orbital polarization of the RIXS and XAS spectra, consistent with experiment, is indicative of low-spin $d^8$ physics for the doped holes.  This is due to the dominance of the unoccupied $x^2-y^2$ weight over that of the $3z^2-r^2$ weight, as reflected in the DFT and DFT+DMFT density of states.

Given the importance of ultraviolet behavior in determining the underlying microscopic Hamiltonian, we hope our results motivate further RIXS studies of both the reduced nickelates \cite{Hepting20,Rossi21,Yao22} as well as cuprates \cite{Lebert17}, particularly at higher loss energies. More generally, we hope the formalism developed here will be useful in simulating RIXS for a broad range of correlated materials.

\begin{acknowledgements} 
J.K., A.J.M.~and M.R.N.~acknowledge funding from the Materials Sciences and Engineering Division, Basic Energy Sciences, Office of Science, US DOE.
Work at the Advanced Photon Source was supported by the U.S.~DOE Office of Science-Basic Energy Sciences, under Contract No.~DE-AC02-06CH11357.
Work at Brookhaven was supported by the U.S.~Department of Energy, Office of Science, Office of Basic Energy Sciences, under Award No.~DE-SC0022311.
H.L.~and A.S.B.~acknowledge support from NSF grant No.~DMR-2045826 and the ASU Research Computing Center for  high-performance computing resources.
The Flatiron Institute is a division of the Simons Foundation.
\end{acknowledgements}

\appendix

\section{Kramers-Heisenberg Formalism}

We start with the Kramers-Heisenberg expression for the RIXS cross section \cite{Mier99,Hariki18} in the dipole approximation
\begin{eqnarray}
\tilde\sigma (E_{in},E_{out},\epsilon,\epsilon^\prime) & = & \sum_f \left|\sum_n\frac{<f|\epsilon^\prime|n><n|\epsilon|i>}{E_n-E_i-E_{in}-i\Gamma_n/2} \right|^2 \nonumber \\
& & \delta(E_{in}-E_{out}-E_f+E_i)
\label{KH}
\end{eqnarray}
where $i, n, f$ are the initial, intermediate and final states, and $\Gamma_n$ is the inverse lifetime of the intermediate
state.  We follow the approach of Ref.~\onlinecite{Mier99} that is illustrated in Fig.~\ref{fig1}.  The RIXS process in the particle-hole approximation involves exciting a $2p$ core electron to an empty $3d$ state, and then having an occupied $3d$ state fill the core hole.  As such, $E_n=E^\prime - E_{2p}$ and $E_f=E^\prime-E$, where $E'$ refers to the unoccupied $3d$ states and $E$ to the occupied ones.  We then use the $\delta$ function to rewrite the energy denominator in Eq.~\ref{KH} as
$E-E_{2p}-E_{out}-i\Gamma/2$ where $\Gamma$ is the core hole broadening.  We now consider the intermediate state sum by writing the matrix elements as:
\begin{eqnarray}
|<n|\epsilon|i>|^2 & = & \sum_{k\sigma}\rho_{k,\sigma}(E^\prime) |<k,\sigma|\epsilon|j>|^2 \nonumber \\
|<f|\epsilon^\prime|n>|^2 & = & \sum_{k\sigma}\rho_{k,\sigma}(E) |<k,\sigma|\epsilon^\prime|j>|^2
\end{eqnarray}
where $j$ are the two $2p$ core states for $L_2$.  We have written the $3d$ states as $|k,\sigma>$ with $k$ = $x^2-y^2$, etc., and $\sigma$ is the spin (the angular and spin decomposition of the $3d$ and $2p$ core states are shown in the main text). Summing over the two $2p$ core states, then from Eq.~\ref{KH} we arrive at the expressions shown in Eqs.~\ref{eq1} and \ref{eq2} of the main text:
\begin{eqnarray}
\tilde\sigma (E_{in},E_{out},\epsilon) & = & \sum_{i,f,\sigma,\sigma^\prime,\epsilon^\prime}
\int dE \rho_{f\sigma^\prime}(E)  \rho_{i\sigma}(E+E_{loss}) \nonumber \\
& &~~~~~~~~~~ \frac{\Gamma}{2}\frac{M_{if\sigma\sigma^\prime}(\epsilon,\epsilon^\prime)}{(E-E_{out})^2+\Gamma^2/4}
\end{eqnarray}
\begin{equation}
M_{if\sigma\sigma^\prime}(\epsilon,\epsilon^\prime) = |\sum_j<f\sigma^\prime|\epsilon^\prime|j><j|\epsilon|i\sigma>|^2
\end{equation}
where $E_{out}=E_{in}-E_{loss}$ and the energy integral ($E$) is over the occupied $3d$ states ($E^\prime=E+E_{loss}$).
When considering this expression, we note that any core hole potential shift in the intermediate state has been ignored, and since the zero for $E_{in}$ and $E$ has been set to the Fermi energy, $E_{out}$ then absorbs $E_{2p}$.  As expected, the integrand in the cross section has a resonance when $E=E_{out}$.
Our expression matches Ref.~\onlinecite{Mier99}, except whereas they approximated the matrix elements by the total $3d$ density of states (their Eqs.~8 and 9), we instead keep the full angular dependence by using the orbitally resolved density of states.  That way, we properly account for the polarization factors associated with the transitions between the $2p$ core states and the $3d$ conduction/valence states.

\bibliography{references}

\end{document}